\documentclass[pmlr]{jmlr}

\usepackage{longtable}

\usepackage{booktabs}

\theorembodyfont{\upshape}
\theoremheaderfont{\scshape}
\theorempostheader{:}
\theoremsep{\newline}

\jmlrvolume{}
\jmlryear{}
\jmlrworkshop{}

\title[The Textbook of Tomorrow]{The Textbook of Tomorrow: Rethinking Course Material Interfacing in the Era of GPT}

 \author{\Name{Audrey Olson} \Email{aolson47@gatech.edu}\\
  \Name{Pratyusha Maiti} \Email{pmaiti6@gatech.edu}\\
  \Name{Ashok Goel} \Email{ashok.goel@cc.gatech.edu}\\
  \addr Georgia Institute of Technology\\
Atlanta, Georgia, USA}

\begin{document}

\maketitle

\begin{abstract}
Online Learning Management Systems (LMSs), such as Blackboard and Canvas, have existed for decades. Yet, course readings, when provided at all, consistently exist as simple digital twins to their real-life counterparts. While online tools and resources exist to help students process digital texts more efficiently or in ways better suited to their learning styles, knowledge about such resources is not evenly distributed and creates a gulf in advantage between students. This paper proposes the courseware integration of ``smart" textbooks, a newfound way for students to chat with their readings, receive summaries and explanations for highlighted text, and generate quiz questions via an AI agent embedded in their online course material. Future iterations of the software aim to add in-context reference highlighting for AI-generated answers and personalized tunings for the end learner.
\end{abstract}

\section{Introduction}
\label{sec:intro}

In his novel \textit{The Diamond Age}, Neil Stephenson introduces the fictional \textit{Young Lady's Illustrated Primer}, a highly intelligent interactive book that propels the story's protagonist to unlikely success following years of tutelage \citep{stephenson}. Among its many features, the Primer is capable of answering the learner's questions, using elements of the learner's environment to enhance teaching via analogy, and providing personalized feedback based off of the learner's actions. With the recent advent of retrieval-augmented generation (RAG) as a means of indexing, searching, and generating answers from document stores, the gap between Stephenson's futuristic vision and the present narrows.

In previous iterations, Jill Watson, a RAG-based virtual teaching assistant employed in online courses \citep{taneja}, operated as a standalone, LMS-integrated chatbot \citep{kakar-jw}. Students could query Jill around course logistics, content clarifications, and any other question that could be answered through the course syllabus and readings. While supplementing some of the more rote work required of human TAs, analysis of agent usage uncovered that a significant number of student questions were concentrated around course material, with a notable presence of high cognitive complexity questions over simple fact finding \citep{maiti}. In light of this discovery, a question arises: how can Jill Watson even better enable deep student engagement with the course material?

In pursuing this question, the following paper introduces a revised interface for Jill Watson not dissimilar to Stephenson's intelligent textbook - a PDF-embedded virtual teaching assistant. We will proceed by discussing the baseline implementation of the project to date, current capabilities, ongoing user studies, and future directions for product optimization.

\section{Implementation and Capabilities}
\label{sec:implementation}

Our intelligent textbook leverages the ReactJS library on top of VTA-GPT, a Python implementation of the RAG pattern which leverages the GPT-4o OpenAI model. At the time of this writing, the GPT-4o OpenAI model yields consistently high performance relative to comparable OpenAI \citep{openai} models and provides the ability to process and generate images and audio in addition to text, features pertinent to future iterations of the interface. Further details around the architecture of VTA-GPT will not be discussed in this paper, given that development of this project existed solely around the client side.

With regards to front-end architecture, the PDF component was made possible through the “react-pdf-viewer” library, which allowed for expedited setup of a full document PDF viewer with minor customizations. The chatbot component of the interface leveraged the open source “react-chatbotify” library. Both choices were made after conducting some high level comparisons of the available open source options, gauging for number of stars and forks on Github, recency of activity, maturity of documentation, ease of setup, and recommendations from the online community. Regarding this last point, for example, the creator of “react-simple-chatbot”, one of the most popular online React-based chatbot libraries with 1.7k stars, promotes the use of the “react-chatbotify” library on his repo’s front-page, given that the “react-simple-chatbot” repository is no longer maintained.

\begin{figure}[htbp]
\floatconts
  {fig:architecture}
  {\caption{Architecture diagram of PDF-embedded Jill Watson. The React Chatbotify library directly connects to VTA-GPT, the GPT-based virtual teaching assistant.}}
  {\includegraphics[width=0.7\linewidth]{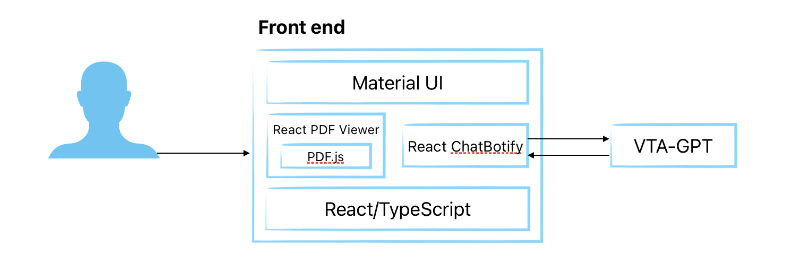}}
\end{figure}

As shown in \textit{Figure 2}, this implementation creates an experience in which students can ``chat" with their readings, empowering them to discover relevant information more quickly, clarify confusing concepts, or engage in a deeper understanding of the text at hand. Through this capability, two of the core components -- text content and AI technologies -- of the pedagogical framework for creating iTextbooks are addressed \citep{ou}, as the provided Q\&A agent leverages the web through GPT-4o as a medium to present content in an effective way. The presence of the Q\&A agent likewise paves a scalable path for applying the learners and assessment blocks of the framework, and GPT-4o's multi-modal capabilities provide ample exploratory room for future visual content through the agent. On top of the capability to ask direct questions, students can highlight long or convoluted passages within the text and receive single click summarizations or simplified explanations delivered through the chatbot interface.  The embedded agent likewise offers different button options, inviting the learner to try generating sample quiz questions and answers as well as the aforementioned features.

\begin{figure}[htbp]
\floatconts
  {fig:interface}
  {\caption{Sample text with embedded Jill Watson}}
  {\includegraphics[width=0.8\linewidth]{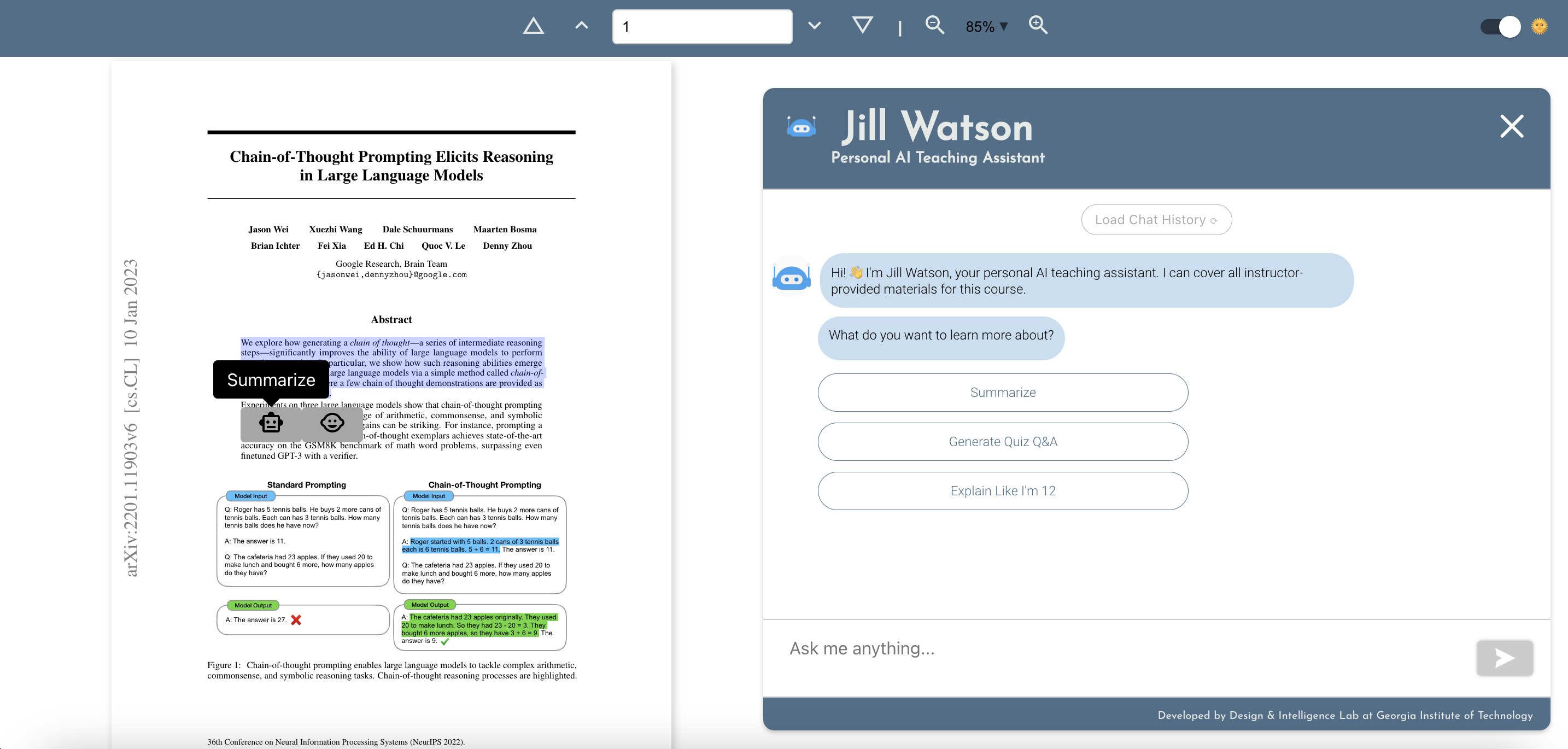}}
\end{figure}

Pending longer term class studies, our expectation is that these features would increase the cognitive presence of Jill Watson, or the abiliity of learners to construct meaning through reflection, discourse, and the application of concepts to real-world situations \citep{lindgren}. For example, should a user ask for a simplification around a specific block of text and receive an analogy for explanation, the user could then proceed to ask further questions around specific components of the analogy or inquire around whether extensions of the analogy could also apply to the original topic at hand. Ideally, this would create a chain of sustained communication that would in turn construct deeper understanding around a particular topic.

\section{User Studies}
\label{sec:userstudies}

Prior to deploying the technology to the virtual classroom, our team has been engaged in conducting an ongoing series of usability studies to collect feedback around system improvement while monitoring for potential learning and engagement gains. These studies leveraged A/B testing in the context of a timed reading comprehension task, providing participants either with the PDF-embedded agent documented here or with the PDF and the AI agent as two separate platforms. Through working with an initial batch of four study participants, two per study option, we were able to gather recurring comments around the slow generation of repetitive and protracted agent responses, a desire for in-context references to the provided answers, and overarching uncertainty around whether the chatbot was specifically trained on the text at hand or more generalized. Other more novel feedback around the AI agent aligned well with Stephenson's vision for personalized teaching based off the requirements of the learner. One suggestion on this front included generating quiz questions via an implementation of the Leitner system, emphasizing more frequent teaching of weaker concepts on a per section basis.

Regarding observed behavior through these studies, user interactions with the interface were surprisingly variable. While one study participant failed to engage with the agent even once through the duration of the study, two participants asked over one dozen questions apiece within the thirty minute time span. In both cases, many questions were follow-ups and clarifications on prior questions, a possible sign of enhanced meaning through discourse. No significant differences in usage were observed between those using the PDF-embedded agent relative to those who used the PDF and the agent as two separate windows or tabs, though the constraints of the study may have encouraged usage in a manner atypical to day-to-day classroom usage of such tools.

We continue to conduct user studies in order to obtain a statistically significant amount of data to determine the PDF-embedded agent's impact on engagement and learning outcomes.

\section{Discussion}
\label{sec:discussion}

\subsection{Addressing User Study Feedback}
\label{sec:userstudyfeedback}

Our study feedback provides ample direction as we proceed with future iterations of our intelligent textbook, as does existing research around multimedia learning. From a usability perspective, our top priority lies in addressing the latency issues experienced when querying the embedded chatbot so that students do not lose trust in the agent's effectiveness to meet their objectives and turn to other AI agents with higher likelihoods of returning misinformation or toxic content. Of the ways to remediate this, the real-time stream of the answer output would be one apparent fix. This interface choice would significantly narrow the learner's gulf of evaluation, or the need to understand the current system state \citep{terry}, when querying the agent. Limiting token count for answer output, whether through API input configuration or through prompt engineering, could likewise ensure shorter, more concise answers that provide more immediate value.

Addressing user uncertainty around the scope of the agent's knowledge, a prioritized improvement would be adding in-context highlighting for the agent's answers. As of today, our Jill Watson endpoint returns both summarized and verbatim versions of the top text references used to generate the response, allowing for keyword matching against the PDF and the opportunity to highlight these reference passages to the learner. In addition to showing the agent's direct tie-ins to the text at hand, this feature likewise ensures groundedness of the provided answer and aids the student learning process by pointing them towards relevant extended reading.

Finally, in addressing the feedback seeking a more custom-tailored learning experience, further envisioning and design alternative proposition would need to be conducted in order to architect an effective personalized agent. One approach, drawing inspiration from Stephen's fictional \textit{Primer}, would be to leverage existing systems such as SAMI (\underline{S}ocial \underline{A}gent
\underline{M}ediated \underline{I}nteractions), a tool used for collecting student demographic data and interests via explicit consent with the intent of connecting them with peers of a shared identity \citep{kakar-sami}. All non-personally identifiable knowledge could then be integrated into the AI agent's system message with the prompt directive to only explain concepts via analogies to the learner's interests. This strategy would apply Mayer's personalization principle of multimedia learning \citep{mayer}, the concept that personalizing content to the learner enables better learning outcomes.

In a different direction on the same topic, integrating the Leitner system into the interface would be a strong improvement upon the current quiz question and answer generation function. By toggling a ``quiz" mode on the AI agent, the learner could proceed to specify a desired reading section before being prompted by the agent around questions unique to the section. Questions answered incorrectly could be cached with metadata indicating a higher priority for revisiting while questions answered correctly would be assigned a lower priority. The likelihood of Jill Watson procuring a novel question would be configured to slowly decrease over time, until only pre-generated questions remained, prompting the user to continue reviewing or restart the process on a new section.

\subsection{Course Integration}
\label{sec:courseintegration}

As the first iteration of the intelligent book is brought to maturity, learning management system (LMS) integration will be crucial in order to ensure the presence of this courseware in the virtual classroom. As of today, the intelligent textbook is capable of displaying a single PDF stored within its codebase. LMS incorporation will require a shift to a multi-PDF model, in which the integrated interface is granted access to the LMS's file store via API in order to allow the display of all course PDFs.

With the ability to monitor intelligent book usage in a classroom environment across multiple semesters and classes, future studies could be conducted around efficacy of the tool in improving long-term learning outcomes and its effect on the Community of Inquiry within the online classroom, or the intersection of cognitive, social, and teaching presence \citep{garrison}. 

\section{Conclusion}
\label{sec:conclusion}
In the spirit of Neal Stephenson's novel \textit{The Diamond Age}, we present an early iteration of the intelligent textbook using the ReactJS front-end library atop a Python implementation of the RAG pattern leveraging GPT-4o. This project introduces one-click summarizations and simplified explanations through highlighting of the text content, a feature intended to increase the cognitive presence of the tool through enhanced enablement of sustained discourse. Through both observed and queried feedback collected through an ongoing series of user studies, we have prioritized a series of future optimizations around real-time answer streaming, highlighting answer references within the PDF itself, and adding student-specific customizations, like teaching in relation to student interests and via the Leitner System. While initial studies of the PDF-embedded agent have reflected indications of increased discourse, future deployments of the tool to a classroom setting will help us examine usage at scale and AI-related effects on learning in a less task and time-constrained environment.

\acks{This work was made possible by NSF Grants \#2112532 and \#2247790 to the National AI Institute for Adult Learning and Online Education. Our thanks is extended to all members of the Design Intelligence Laboratory for their input into this work.}

\bibliography{pmlr-sample}

\end{document}